# An end-to-end framework for gene expression classification by integrating a background knowledge graph: application to cancer prognosis prediction

Kazuma Inoue[1], Ryosuke Kojima[1, *], Mayumi Kamada[1], Yasushi Okuno[1,2, *]

[1]Graduate school of medicine, Kyoto University, Sakyo-ku Kyoto, Kyoto 606-8507, Japan

[2]RIKEN Center for Computational Science, Kobe, Hyogo 650-0047, Japan

*To whom correspondence should be addressed.

## Abstract

**Motivation:** Biological data may be separated into primary data, such as gene expression, and secondary data, such as pathways and protein-protein interactions. Methods using secondary data to enhance the analysis of primary data are promising, because secondary data have background information that is not included in primary data. In this study, we proposed an end-to-end framework to integrally handle secondary data to construct a classification model for primary data. We applied this framework to cancer prognosis prediction using gene expression data and a biological network.

**Results:** Cross-validation results indicated that our model achieved higher accuracy compared with a deep neural network model without background biological network information. Experiments conducted in patient groups by cancer type showed improvement in ROC-area under the curve for many groups. Visualizations of high accuracy cancer types identified contributing genes and pathways by enrichment analysis. Known biomarkers and novel biomarker candidates were identified through these experiments.

**Availability:** This framework is available at https://github.com/clinfo/SLGCN_cancer_prognosis.

**Contact:** kojima.ryosuke.8e@kyoto-u.ac.jp, okuno.yasushi.4c@kyoto-u.ac.jp

**Supplementary information:** Supplementary data are available at *Bioinformatics* online.

## 1 Introduction

Biological systems are supported by interactions between various molecules that conduct biological activities (Aloy, P. and Russell, R., 2008; Braun, P. and Gingras, A.C., 2012). To comprehensively understand biological systems, integrating various types of data, such as experimental omics and the literature, is important. In general, such data can be categorized into primary and secondary data. The former may be defined as data directly measured from experiments, such as gene expression data, obtained from a patient sample. The latter refers to data obtained by analyzing and aggregating primary data, such as signal transduction pathways and protein-protein interactions (PPIs). The secondary data obtained by analyzing primary data is disseminated through publications and databases, which may be reused as background knowledge for further research. Studies related to primary data analysis focus on differences in genomic variants and gene expression among samples. Recently, the use of deep learning models for such primary data has been reported (Issa, N.T. et al., 2021). For example, prognosis prediction from gene expression data using deep learning was reported (Poirion, O.B. et al., 2021). In contrast, advanced analysis of secondary data often focuses on molecular networks, such as pathways and PPIs. It addresses issues, such as link prediction, subgraph extraction, and addresses issues, such as link prediction, subgraph extraction, and topology classification (Abdel-Hafiz, M. et al., 2022; Bamunu Mudiyanselage, T. et al., 2022). With recent developments in deep learning, graph neural networks (GNN) have achieved a state-of-the-art for many tasks related to the analysis of secondary data (Defferrard, M. et al., 2016; Hamilton, W. et al., 2017; Kipf, T.N. and Welling, M., 2017).

Approaches to complement secondary data using primary data have also been reported. Examples include methods to obtain subnetworks related to breast cancer metastasis using gene expression data and detect important nodes on the graph representing molecular networks using microarray data (Chuang, H.Y. et al., 2007; Emig, D. et al., 2013). In contrast, approaches that use secondary data to enhance primary data analysis exist (Chereda, H. et al., 2019; 2021; Ramirez, R. et al., 2020; 2021). For example, using cancer pathway as secondary data, cancer prognosis was predicted from gene expression data. (Zheng, X. et al., 2020; 2021). The use of secondary data to enhance primary data analysis is a promising approach, as it enables one to leverage the vast amount of background information that is not included in the primary data. However, previous methods used separately processed secondary data as additional features of the primary data; thus, these methods lacked an end-to-end framework to efficiently integrate these data types.

We propose a novel deep learning framework to integrally analyze secondary data, such as a molecular network, to construct a prediction model from primary data, such as omics data. This enables a prediction at individual levels using background information of the biological network. We apply the framework to cancer prognosis prediction using gene expression data and a biomolecular network.

Prognosis in cancer varies considerably from patient to patient, in part, because of genetic differences. Therefore, various deep learning methods to predict cancer patient survival based on genetic information have been developed (Ching, T. et al., 2018; Huang, Z. et al., 2020; Katzman, J.L. et al., 2018; Pavageau, M. et al., 2021). In this study, we applied our framework to the problem of predicting prognosis considering the background network. Specifically, our proposed framework consists of two parts: a GNN, which trains molecular interactions represented by a graph as background knowledge, and a deep neural network (DNN), which predicts patient survival based on gene expression data that varies for each individual. This model predicts patient survival over specific years by utilizing biomolecular interaction information and gene expression data from an individual patient. We evaluated our model using The Cancer Genome Atlas (TCGA) database and confirmed the effectiveness of the proposed framework by comparison with conventional models.

# 2 Methods

## 2.1 Knowledge graph construction

In this study, a knowledge graph representing molecular interaction information was constructed using the Pathway Commons database (Cerami, E.G. et al., 2011), which is a large-scale dataset containing biological pathways and interactions in various datasets. It primarily consists of two biomolecules and their relationship. The knowledge graph consists of nodes and edges representing biomolecules and the relationship between them. There were 13 types, such as "phosphorylation" and "chemical affects."

## 2.2 Individual cancer patient information

Individual cancer patient information for 33 different cancers were obtained from TCGA. The gene expression data for cancer patients estimated from RNA-Seq was acquired by recount2 (Collado-Torres, L. et al., 2017) and log-transformed from transcripts per million for features using the natural logarithm. Recounts2 is an online resource that compiles gene expression data obtained from many research projects, including TCGA. Here, we selected only cancer-related genes. The genes listed for the Molecular Signature Database (MsigDB) (Liberzon, A. et al., 2015) and the LM22 immune gene signatures (Newman, A.M. et al., 2015) were selected. Genes that did not exist in the knowledge graph nodes were eliminated. The expression data for 4,448 genes were used. We also obtained clinical information for each patient, such as overall survival and cancer type, which was summarized by Liu J et al (Liu, J. et al., 2018). The cancer types were encoded into 33-dimensional one-hot vectors and used as features for each patient.

## 2.3 Sample selection and labeling

Verified samples were selected. Patients were divided based on a median number of censored days (819 days). Samples with prolonged survival were considered to have responded to treatment with drugs or surgery. We excluded 364 samples who survived over 3,595 days, which corresponded to the top 5% in order of survival time of censored patients with more than 819 days as well as deceased patients. The remaining 10,823 samples were labeled as alive or dead within verified years. We predicted 1- to 5-year patient survival. For the "$n$" year prediction ($n$ = [1, 2, …, 5]), the samples censored within "$n$" years ($n$ * 365 days) were excluded because the survival states when "$n$" years passed could not be judged. If patients survived over "n" years, a label "1" was assigned, and if they did not, a label "0" was assigned. Samples details are shown in Supplementary Table 1.

## 2.4 Model architecture

Our proposed framework consisted of two parts: the GNN for calculating molecular interaction features and the DNN for predicting patient prognosis using gene expression data.

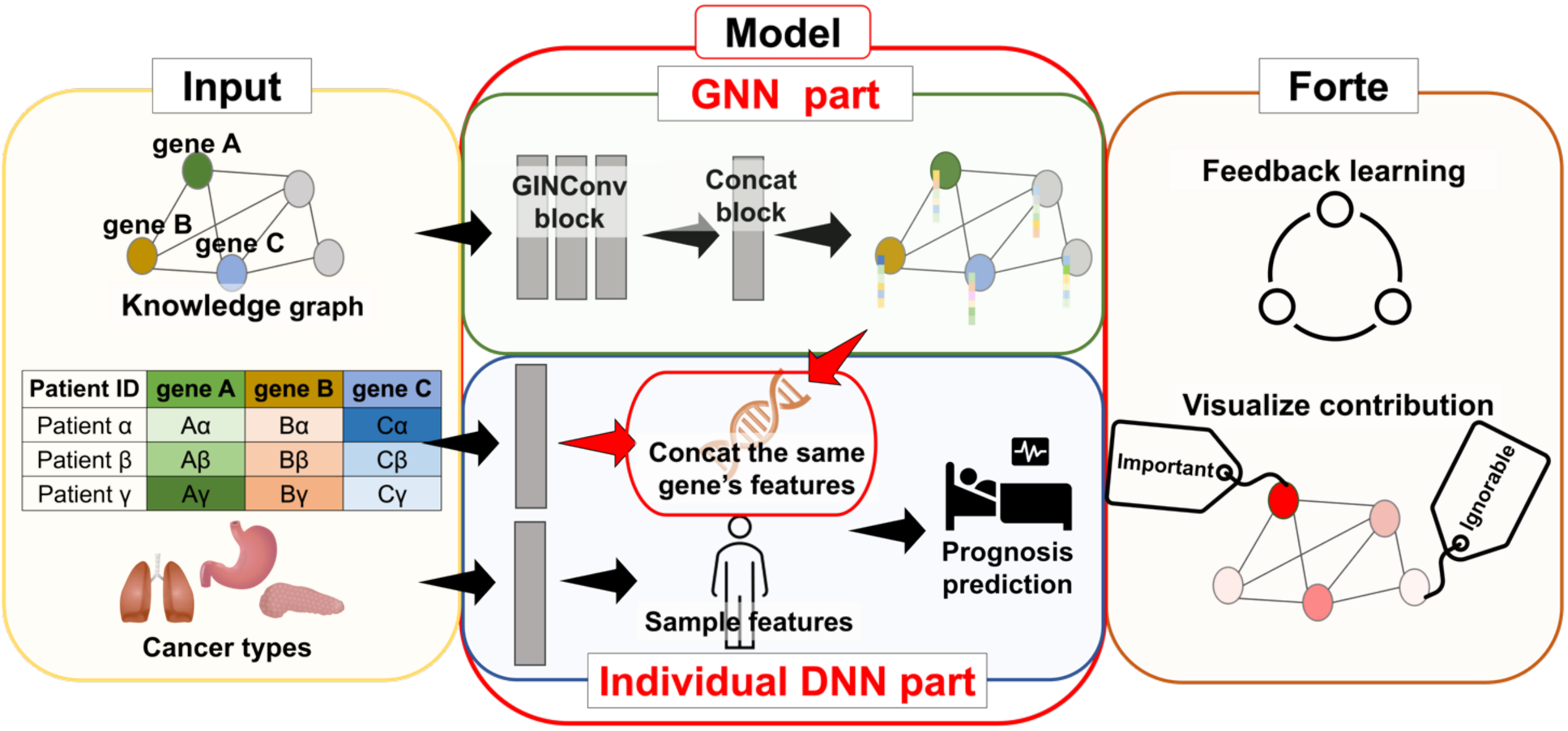

**Fig. 1. Model architecture.** There are three inputs: a knowledge graph, patient gene expression data, and cancer type. The model has two parts: the GNN part and the NN part. Prognoses are predicted in the DNN part. Feedback learning and visualization of feature contribution can be conducted.

### 2.4.1 GNN part

In the GNN part, the knowledge graph representing molecular interaction information was used as the input and latent vectors for each graph node were obtained from the training step. The knowledge graph $G$ represents the molecular interaction as background knowledge and is described as $G = (V, E)$, in which $V$ and $E$ are finite sets of nodes and edges, respectively. Here, $E$ is described as $E = (u, r, v)$: $u$ and $v$ are graph nodes and $r$ is an edge between them. The GNN part consisted of three layers of a Graph Isomorphism Network (GIN) (Xu, K. et al., 2019), Conv Block and a Concatenate Block. The latent vector $\boldsymbol{z}_u$ of each node $u$ was calculated as follows.

$$\boldsymbol{z}_u = GNN(u, G) \quad (1)$$

The calculation formula for the latent vectors in the GIN layer was defined by formula (2).

$$Z^{\ell+1} = \sigma\left(\sum_r W(r) \cdot \sigma\left(GINConv\left(Z^{\ell}, G_{(r)}\right)\right) + b_{(r)}\right) \quad (2)$$

$Z^{\ell}$ represents a matrix consisting of the node vectors in the $\ell$ layer and $G_{(r)}$ is a subgraph having only edges meaning relation $r$.

### 2.4.2 Individual DNN part

For the DNN part, cancer patient prognosis was predicted using the latent vectors calculated in the GNN part and the clinical information for each patient. The input dataset $D$ was defined as $D = \{(\boldsymbol{X}_1, y_1), (\boldsymbol{X}_2, y_2), \dots, (\boldsymbol{X}_N, y_N)\}$. Here, $\boldsymbol{X}_i = (\boldsymbol{u}, \boldsymbol{f}_u)$, $\boldsymbol{u}$ is a gene, and $\boldsymbol{f}_u$ is the expression value for gene $\boldsymbol{u}$ of patient $i$. $y_i$ is a survival label, and if sample $i$ died within "$n$" years, $y_i = 0$, otherwise $y_i = 1$.
The output in the DNN part was as follows.

$$\hat{y}_i = NN(\mathbf{X}_i, \{\boldsymbol{z}_u : u \in V\}) \quad (3)$$

The DNN part consisted of the two-modal block, the aggregation block, and the multi-modal block. First, the two-modal block applied to all genes having expression features and using $z_u$ and gene expression features $\boldsymbol{f}_u$. The output was defined as follows:

$$s_u = \boldsymbol{z}_u + \mathbf{W}\boldsymbol{f}_u \quad (4)$$

$W$ was a parameter matrix of this neural network.

Second, the output in the aggregation block was calculated using $s_u$,

$$t = h\left(\sum_u g(s_u)\right) \quad (5)$$

in which $g$ and $h$ were computed as the multi-layer perceptron (Zaheer, M. et al, 2017). An activation function in the middle layers was Exponential Linear Unit and batch normalization was introduced. A prediction value $\hat{y}$ was calculated using $t$ and sample vectors $\boldsymbol{e}_i$. Sample vectors represent patient clinical information. In this study, we included their cancer types; thus, there were 33-dimensional one-hot vectors.

The correct labels were 0 or 1, so the proposed framework was a binary classification. The activate function in the output layer is a sigmoid function, so the prediction value $\hat{y}$ was a real number between 0 to 1. We conducted a 5-fold cross-validation for each verification year for accuracy evaluations.

$$\hat{y}_i = Sigmoid(MNN(t, \boldsymbol{e}_i)) \quad (6)$$

## 2.5 Learning methods

Model learning was conducted in two steps. In the first step, the link prediction was adapted with the graph to pre-train the GNN. Then, fine-tuning for the whole network including the DNN part was performed. We could select whether these two steps were connected or not. When they are connected, it is designated end-to-end. Its fine-tuning is conducted throughout the entire model. Link prediction is a learning method for the probability of existing an edge between two nodes. In other words, it predicts whether an edge exists between two nodes. By pre-training the GNN part with link prediction, the graph structure represents biomolecular interaction information. The loss function of the link prediction was defined as follows:

$$L_{pre} = -\ell_{+}(u, v) - \ell_{-}(u, v') \quad (7)$$

$u$ and $v$ were sampled from $E$ at random and $v'$ was sampled from $V$ at random.

$$\ell_{+}(u, v) = log(\sigma(u^T v)) \quad (8.1)$$

$$\ell_{-}(u, v) = log(\sigma(-u^T v)) \quad (8.2)$$

When a link with relation was predicted, a weight matrix **W** was used and the Dist-Mult model (Yang, B. et al., 2014) was employed, which is a widely used method for knowledge graph embedding to complement a relation between the nodes.

$$\ell_{+}(u, r, v) = log(\sigma(u^T \mathbf{W}_r v)) \quad (9.1)$$

$$\ell_{-}(u, r, v) = log(\sigma(-u^T \mathbf{W}_r v)) \quad (9.2)$$

Fine-tuning was done by backward computation using a loss function of $\hat{y}$ and $y$. In the DNN part, a sigmoid function was employed as an activation function in the output layer and binary cross-entropy was used as a loss function.

$$L_{fine}(\hat{y}, y) = -\hat{y} log y - (1 - y) log\,(1 - \hat{y}) \quad (10)$$

### 2.6 Model evaluation

In this study, 1- to 5-year patient survival predictions were performed and we evaluated these prediction models by a 5-fold cross-validation. Using the final epoch in the training set of each fold, the area under the curve (AUC) of the test set was calculated and mean AUC values were used for the evaluation. We used scikit-learn v0.24.2 (Pedregosa, F. et al., 2011) and pytorch v1.7.0 (Paszke, A. et al., 2019) for model implementation and evaluation. Our framework had two modes: the end-to-end or not. The end-to-end mode learns through the GNN part and the DNN part at once. The not end-to-end model conducts pre-training of the GNN part and updates only the DNN part weights. In other words, the end-to-end model can update graph node features depending on the DNN prediction.

### 2.7 Prediction interpretation

To determine which features contribute to survival prediction, we used the integrated gradients (IG) (Sundararajan, M. et al., 2017) method to visualize the features contribution. The IG of the patient $i$ was defined by using baseline $x'$ and input features $x$. A baseline $x'$ is a standard value when the model determines their contributions and it is used when their feature values are 0. Thus, the IG of a patient $i$ is defined as follows:

$$IG_i(x) = (x_i - x'_i) \int_{\alpha=0}^{1} \frac{\partial F(x' + \alpha(x - x'))}{\partial x_i} d\alpha \quad (11)$$

The greater the positive IG values, the more this feature contributes to patient survival and the greater the negative IG values, the more it contributes to death.

The evaluation method differs from the graph node IG and gene expression IG. The latter is calculated by formula (11) and evaluates the contribution to prognosis. IG values for the GNN part are calculated for each dimension of the latent vector ($z_u$) for each node. That is, the IG value of each node in the GNN part has the same dimensions as the latent vector (the number of dimensions $C$ is 32 in this study). As a result, positive and negative IG values may be mixed in each node and it may be difficult to interpret whether the node contributes to death or survival. Therefore, we normalized the IG value for each node using the L2 norm and the normalized values were used to determine the contribution of each node for prediction. The IG value for gene $j$ in the patient $i$ was calculated as follows:

$$IG_{ij} = \frac{\left\| {x_{ijn}}^2 \right\|_2}{\sum_{k=1}^{n} IG_{ik}} \quad (12)$$

where, $x_{ijn}$ represents the $n$-th dimensional $(0 \leq n \leq C)$ IG. $\| \|_2$ is the L2 norm.

We conducted 5-fold cross-validation with random seeds and calculated the average IG values. The contribution evaluation was performed with the average IG values and the results were visualized with Cytoscape (Shannon, P. et al., 2003), which is an open-source software used to visualize complex networks.

## 3 Results

### 3.1 Overall accuracy for all years

We constructed a knowledge graph for input in the GNN part using the Pathway Commons datasets. The knowledge graph consisted of 30,919 nodes, 1,884,849 edges, and 13 edge types.

To verify the effect of the proposed framework consisting the GNN and DNN parts, it was compared with the model using only the DNN part, which corresponded to the conventional model using patient profiles. Differences in the accuracy of the model with and without cancer type data for each patient were also examined. The ROC-AUC values for each model are shown in Table 1.

**Table 1.** ROC-AUC of four models for all verification years. The column "year" represents the verification year. Models predicting whether patients died within each verification year.

| year | DNN | DNN + cancer types | GNN + DNN | GNN + DNN + cancer types |
|---|---|---|---|---|
| 1 | 0.6312 | 0.7425 | 0.7382 | 0.7585 |
| 2 | 0.6168 | 0.7672 | 0.7678 | 0.7596 |
| 3 | 0.6188 | 0.7624 | 0.7733 | 0.7890 |
| 4 | 0.6173 | 0.7674 | 0.7714 | 0.7900 |
| 5 | 0.5971 | 0.7644 | 0.7581 | 0.7850 |

We selected the end-to-end model, in which pre-training was 50 epochs, and analyzed the results. For all verification years, our proposed framework (GNN + DNN part) outperformed the conventional model (only the DNN part). It is important to note that our model (GNN + DNN) achieved equivalent accuracy to the model with the DNN part and cancer types. In addition, when cancer types were added to our model, the performance yielded the highest accuracy.

Fig. 2 shows how the ROC-AUC for the 5-year models changed by pre-training epochs. Our model was selected by either adopting the end-to-end method or not. It can learn consistently two parts if selecting the former, and if not, it cannot renew the GNN part and only learn from the DNN part. In addition, we selected pre-training epochs arbitrarily and found that it could predict more highly and stably if "end-to-end" was selected for the learning method.

## 3.2 Accuracy differences by cancer type

In this section, we considered how the accuracy of each cancer type changed with or without the GNN part. Fig. 3. shows the accuracy comparison of the two models for the 3-year prediction. The total number of cancer types used for learning and prediction was 33. In this experiment, some of the AUCs could not be calculated because of the number of patients that survived and died in each fold on cross-validation. Fig. 3 shows that our model resulted in higher ROC-AUC values for many cancer types compared with the conventional model. For all verification years, 83–95% of the cancer types had improved ROC-AUC values compared with the conventional model. In particular, adrenocortical carcinoma (ACC), kidney renal papillary cell carcinoma (KIRP), brain lower grade glioma (LGG), and mesothelioma (MESO) had more accurate predictions compared with the others. We confirmed this tendency for all verification years. Although we found no correlation between accuracy improvement and the corresponding patient death ratios and sample sizes, specific associations with clinical features and onset organs were not observed.

## 3.3 Visualization of the feature contributions using integrated gradients

We used the IG method to examine the contributed inputs to the prediction, including nodes in the GNN part and features in the DNN part.

### 3.3.1 IG analysis: graph nodes

We calculated the IG values for the graph nodes in the GNN part and determined the relationship between the characteristics of the graph nodes and the IG values. Centrality measures the degree to which a graph node is central and acts as a hub in a knowledge graph. The degree of centrality represents the number of edges that a node contains. Fig. 4 shows a positive correlation (R = 0.727) between the IG value and the degree of centralities for the graph nodes. Other centrality measures, such as the closeness of centrality, also exhibited the same tendency (Details in Supplementary Fig. 1). It showed that the nodes with a high IG value were high centrality nodes in the knowledge graph. Nodes with high centrality are important for the interaction with various molecules. We confirmed that these crucial nodes contributed to prediction.

We performed an enrichment analysis using the top 100/300/500/1000 average IG values for the graph nodes using the Database for Annotation, Visualization, and Integrated Discovery (DAVID) (Huang, D.W. et al., 2007) as well as gene sets from the Kyoto Encyclopedia of Genes and Genomes (KEGG) database (Kanehisa, M. and Goto, S. 2000; Kanehisa, M. et al., 2016) (Table 2 and Supplementary Table 2). Table 2 shows the result for the top 10 pathways for the top 100 nodes, in which cancer-associated pathways were identified. This indicates that our model learned that the nodes representing cancer-associated biomolecules in the graph contributed to prediction.

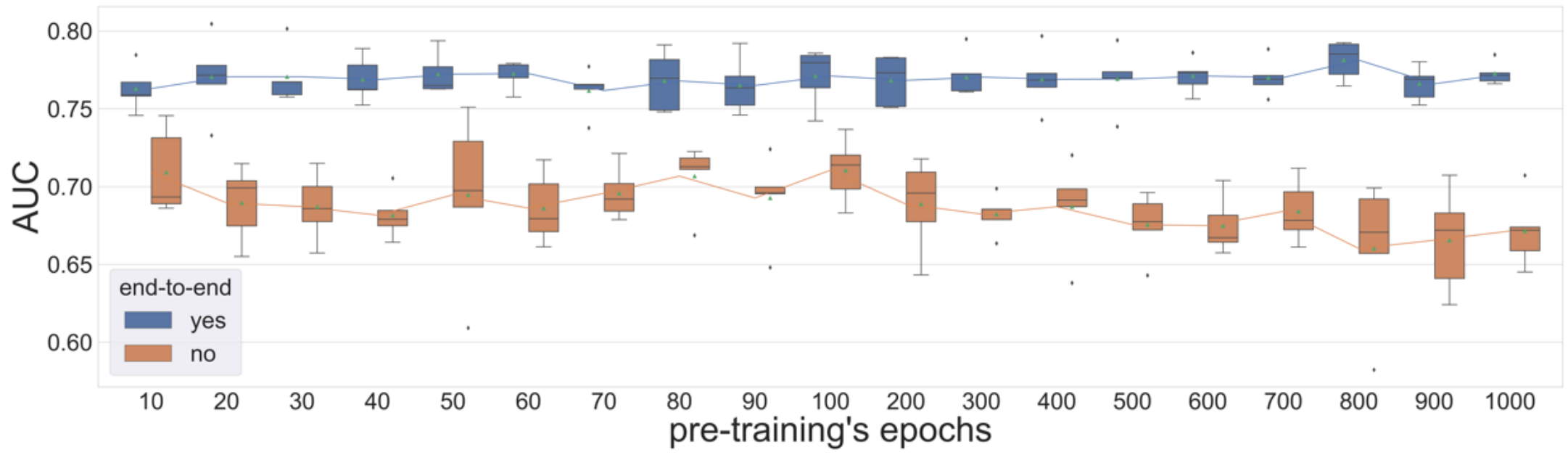

**Fig. 2. Accuracy transitions in the 5-year model with or without pre-training in the GNN part.** The vertical axis and horizontal axis represent the AUC and pre-training epochs, respectively. Blue boxes represent the end-to-end model AUC and orange boxes are not the end-to-end model.

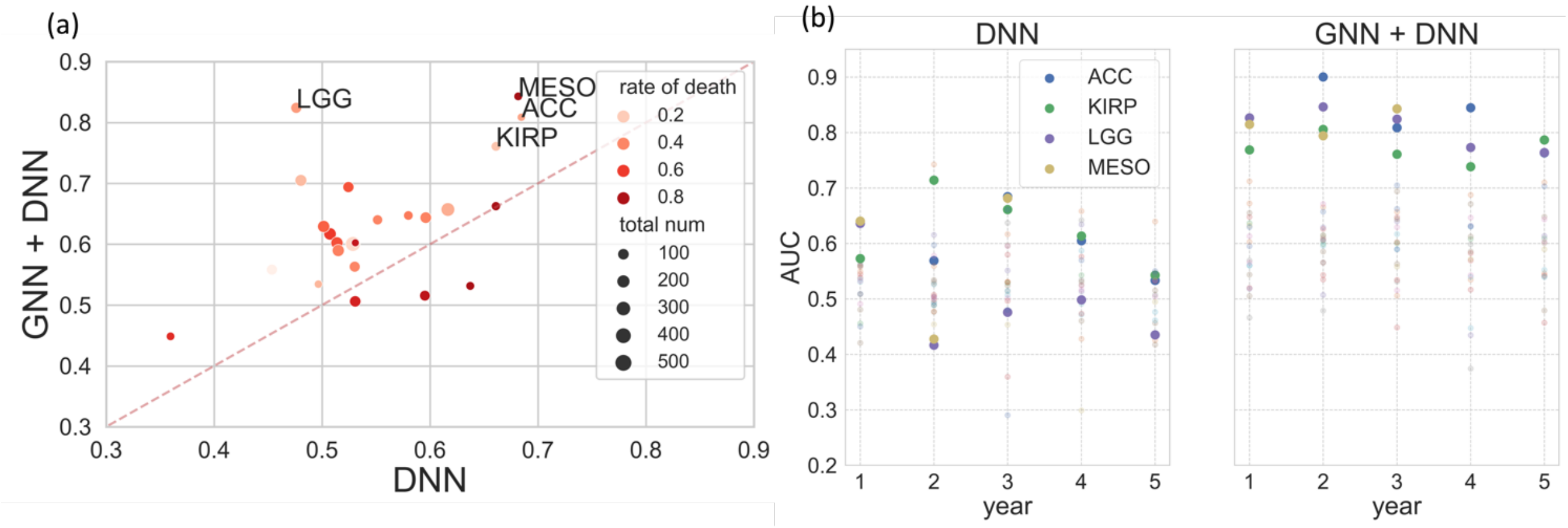

**Fig. 3.** a) The scatter plot of each cancer's ROC-AUC in 3 year's model. The vertical axis and horizontal axis represent the GNN + DNN model's AUC and only the DNN part model's, respectively. Each dot color represents the death ratio and each dot size does its sample size of each cancer. b) The AUC transition of all cancer types for 5 years.

### 3.3.2 IG analysis: graph nodes for individual cancer types

For all 33 cancer types, we made node lists, in which the IG values were ranked in the top 1500. The top 1500 nodes of IG values were obtained and compared for differences among all cancer types. Focusing on the four cancer types (ACC, KIRP, LGG, and MESO) for which the addition of the GNN part significantly improved accuracy in section 3.2, we examined the effect of the graph in detail.

**Table 2.** The enrichment analysis result for the Top 100 IG nodes.

| Rank | KEGG Term | Count | P-value |
|---|---|---|---|
| 1 | Pathways in cancer | 49 | 4.61E-29 |
| 2 | Kaposi sarcoma-associated herpesvirus infection | 33 | 4.48E-27 |
| 3 | Hepatitis B | 31 | 6.41E-27 |
| 4 | Lipid and atherosclerosis | 32 | 2.31E-24 |
| 5 | Prostate cancer | 24 | 2.71E-23 |
| 6 | Human cytomegalovirus infection | 31 | 1.52E-22 |
| 7 | AGE-RAGE signaling pathway in diabetic complications | 22 | 4.05E-20 |
| 8 | Chemical carcinogenesis – receptor activation | 28 | 9.48E-20 |
| 9 | Proteoglycans in cancer | 27 | 5.60E-19 |
| 10 | Endocrine resistance | 21 | 6.29E-19 |

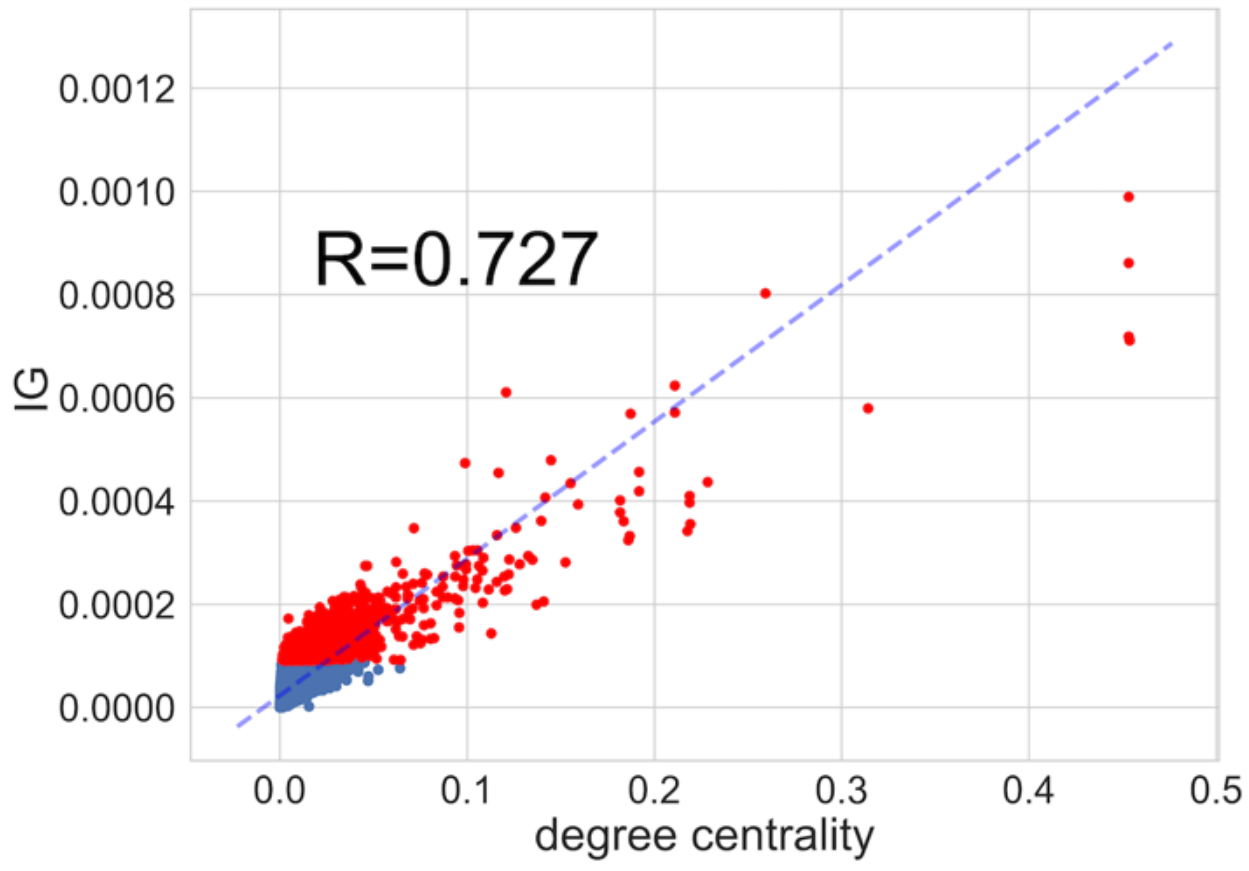

**Fig. 4. The graph node IG scatter plot.** The vertical axis and horizontal axis represent IG values and the degree of centrality for each node in the graph, respectively. The red dots represent the nodes ranking in the top 1500 and the blue dots represent the others.

We performed a t-test for each listed node for one caner type and the other 32 cancer types as well as for the nodes exhibiting high IG values (p-value < 0.05). For KIRP, IG values for the 17 nodes in Fig. 5 were significantly different from that of the other cancer types. These nodes were connected in the input graph. CHEBI:2504 indicates an aflatoxin node that acts a hub node. Aflatoxin is a carcinogen, particularly in the liver, and its relevance to kidney cancer is known (Bbosa, G.S. et al., 2013; Li, H. et al., 2018; Marchese, S. et al., 2018). In addition, 56 nodes showed significant differences for MESO (Supplementary Table 3). For these nodes, we conducted an enrichment analysis using the KEGG pathway database. The results indicated that the nodes were enriched in the Mitogen-activated Protein Kinase (MAPK) signaling pathway. A series of events (FGF2 induction, MAPK pathway activation, and MMP1 induction) are known to be important for epithelial-to-mesenchymal transition (EMT) signaling in MESO (Schelch, K. et al., 2018; Ramundo, V. et al., 2021). For ACC, six nodes were identified as significantly different nodes (Supplementary Table 3). SF1 is known as a diagnostic marker for ACC (Almeida, M.Q. et al., 2010; Ehrlund, A. et al., 2012).

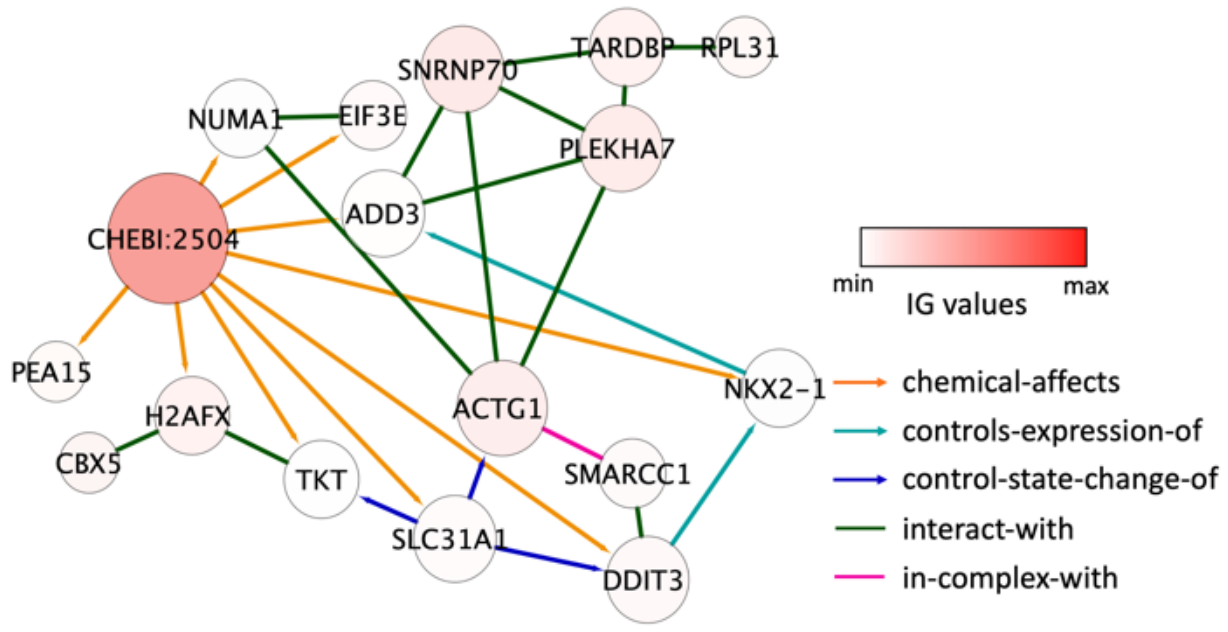

**Fig. 5. The KIRP subgraph.** The nodes exhibiting significant differences in a t-test. The colors represent IG values and their sizes indicate their degree centralities.

### 3.3.3 IG analysis; the gene expression

The top 200 gene lists for each cancer were obtained by calculating the IG values for the gene expression data. The top 200 genes consisted of the top 100 death-contributing genes and the top 100 survival-associated genes. For the top 200 genes, we identified differences among the models and cancer types as well as between genes with cancer type-specific IG values.

We compared these gene lists among the three models (DNN, DNN + cancer types, GNN + DNN). Fig. 6 shows a Venn diagram comparing the three model gene lists for LGG. The genes were altered when the GNN part was added and this tendency was confirmed in almost all cancer types. It revealed that gene expression features were changed and contained genes that were associated with prognosis, whereas the accuracy improved because of the molecular interaction information. We confirmed that the ROC-AUC was improved by the GNN part in the previous section (Table 1). The genes listed only in the GNN + DNN part may contribute to high accuracy prediction.

Next, we investigated cancer-specific relevant genes by comparing the gene lists of the GNN + DNN model among the various cancer types. FURIN, PLAC8, PBK, and LMNB1 were listed only for LGG. FURIN and PBK are known prognostic factors and their increased expression is associated with poor prognosis in LGG (Feng, T. et al., 2021; Zhou, B. and Gao, S. 2021). These results indicate that genes uniquely listed in each cancer type are related to prognosis.

For the top 200 IG genes of the GNN + DNN model, we obtained genes with IG values significantly (p-value $< 0.05$) different from other cancer types. In MESO, the IG values of ITGA10 and COL4A1 were significantly high. These two genes are related to an extracellular matrix (ECM) receptor, which is a non-cellular element and contributes to tissue morphogenesis and differentiation. ECM is associated with the growth of MESO cells and is considered a potential treatment target (Pass, H.I. et al., 2005; Tajima, K. et al., 2010). In LGG, 81 genes exhibited significantly high IG values (Supplementary Table 4). An enrichment analysis was performed for these genes (Table 3) and several were associated with cranial nerve diseases, such as spinocerebellar degeneration, prion diseases, and Alzheimer's disease. These results indicate that our model utilized features clinically relevant to cancer prognosis and clinical conditions for each cancer type. In addition, the results suggest that the graph nodes and genes contributing to prediction may represent novel biomarkers.

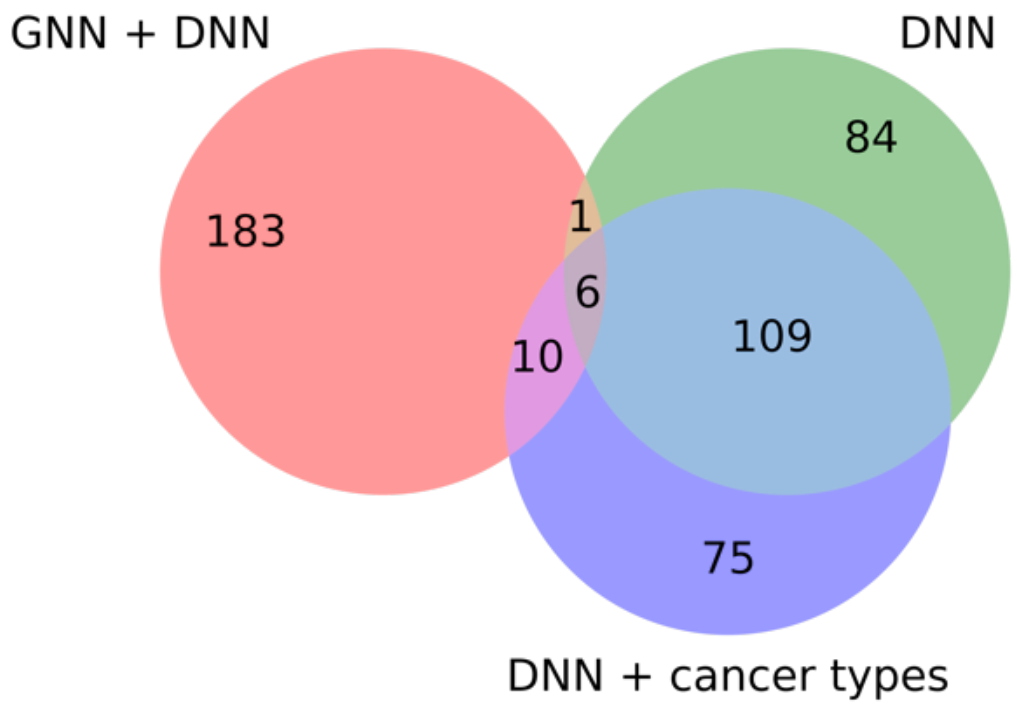

**Fig. 6. A Venn diagram comparing the top 200 genes of the three models.** The red circle represents the GNN + DNN part, the green represents the DNN part, and the blue represents the DNN + cancer types.

**Table 3.** The enrichment analysis results for the 81 high IG value genes for LGG.

| Rank | KEGG Term | count | P-value |
|---|---|---|---|
| 1 | Prion disease | 7 | 2.17E-3 |
| 2 | Pathways of neurodegeneration – multiple diseases | 8 | 8.52E-3 |
| 3 | Parkinson disease | 6 | 1.00E-2 |
| 4 | Alzheimer disease | 7 | 1.13E-2 |
| 5 | Huntington disease | 6 | 1.74E-2 |
| 6 | Proteasome | 3 | 2.19E-2 |
| 7 | Amyotrophic lateral sclerosis | 6 | 3.38E-2 |
| 8 | Spinocerebellar sclerosis | 4 | 3.41E-2 |
| 9 | Epstein-Barr virus infection | 4 | 7.93E-2 |
| 10 | Glycosaminoglycan biosynthesis – chondroitin sulfate / dermatan sulfate | 2 | 9.54E-2 |

## 4 Conclusion

We propose a new end-to-end framework integrating primary data and secondary data represented as a graph and applied this framework to predict cancer patient prognosis. We regarded biomolecular interactions as a background knowledge graph and individual gene expression data as primary data. Compared with the conventional prediction model for cancer patient prognosis, we improved prediction accuracy by combining the biomolecular information and the individual gene expression data. Moreover, the IG method enabled us to visualize which nodes represented genes and genes that had prognostic value.

Our results showed graph nodes and gene expression data with high IG values, contributed inputs, and were consistent with known biological information, such as cancer-related pathways and prognostic factors. We obtained new genes with high IG values, which may potentially represent novel biomarker candidates. In addition, our model without cancer type labels could achieve a similar prediction accuracy. The results suggest that our model captures potential biological knowledge for the target disease, thus our framework may also be useful for other diseases.

To apply these findings to the clinic, more accuracy is needed. Rich clinical information improves the prediction accuracy (Huang, S. et al., 2014). We used only cancer types as patient information. To improve accuracy for clinical applications, additional clinical information, such as cancer stage, sex, age, and medical history is needed.

## Funding

This work was supported by MEXT as "Program for Promoting Researches on the Supercomputer Fugaku" (Application of Molecular Dynamics Simulation to Precision Medicine Using Big Data Integration System for Drug Discovery, JPMXP1020200201) and Cabinet Office, Government of Japan, Public/Private R&D Investment Strategic Expansion Program（PRISM）. This research was also supported by JST Moonshot R&D Grant Number JPMJMS2024, and JSPS KAKENHI Grant No.21H03537, Japan.

## References

Abdel-Hafiz, M. *et al.* (2022) Significant subgraph detection in multi-omics networks for disease pathway identification. *Front. Big Data*, **5**, 894632.

Almeida, M.Q. *et al.* (2010) Steroidogenic factor 1 overexpression and gene amplification are more frequent in adrenocortical tumors from children than from adults. *J. Clin. Endocrinol. Metab.*, **95**, 1458–1462.

Aloy, P. and Russell, R. (2008) Targeting and tinkering with interaction networks. *FEBS Lett.*, **582**, 1219.

Bamunu Mudiyanselage, T. *et al.* (2022) Predicting CircRNA disease associations using novel node classification and link prediction models on Graph Convolutional Networks. *Methods*, **198**, 32–44.

Bbosa, G.S. *et al.* (2013-01-23) Review of the biological and health effects of aflatoxins on body organs and body systems. *Aflatoxins*.

Braun, P. and Gingras, A.C. (2012) History of protein-protein interactions: from egg-white to complex networks. *Proteomics*, **12**, 1478–1498.

Cerami, E.G. *et al.* (2011) Pathway Commons, a web resource for biological pathway data. *Nucleic Acids Res.*, **39**, D685–D690.

Chereda, H. *et al.* (2019) Utilizing molecular network information via graph convolutional neural networks to predict metastatic event in breast cancer. *Stud. Health Technol. Inform.*, **267**, 181–186.

Chereda, H. *et al.* (2021) Explaining decisions of graph convolutional neural networks: patient-specific molecular subnetworks responsible for metastasis prediction in breast cancer. *Genome Med.*, **13**, 42.

Ching, T., Zhu, X. and Garmire, L.X. (2018) Cox-nnet: an artificial neural network method for prognosis prediction of high-throughput omics data. *PLOS Comput. Biol.*, **14**, e1006076.

Chuang, H.Y. *et al.* (2007) Network-based classification of breast cancer metastasis. *Mol. Syst. Biol.*, **3**, 140.

Collado-Torres, L. *et al.* (2017) Reproducible RNA-seq analysis using recount2. *Nat. Biotechnol.*, **35**, 319–321.

Defferrard, M. *et al* Convolutional neural networks on graphs with fast localized spectral filtering. In, Proceedings of the 30th international conference on neural information processing systems. Barcelona, Spain (2016). p. 3844–3852.

Ehrlund, A. *et al.* (2012) Knockdown of SF-1 and RNF31 affects components of steroidogenesis, TGFβ, and Wnt/β-catenin signaling in adrenocortical carcinoma cells. *PLOS ONE*, **7**, e32080.

Emig, D. *et al.* (2013) Drug target prediction and repositioning using an integrated network-based approach. *PLOS ONE*, **8**, e60618.

Feng, T. *et al.* (2021) PDZ binding kinase/T-LAK cell-derived protein kinase plays an oncogenic role and promotes immune escape in human tumors. *J. Oncol.*, **2021**, 8892479.

Hamilton, W. *et al.* (2017) Inductive representation learning on large graphs. *Adv. Neural Inf. Process. Syst.*, 1024–1034.

Huang, D.W. *et al.* (2007) DAVID Bioinformatics Resources: expanded annotation database and novel algorithms to better extract biology from large gene lists. *Nucleic Acids Res.*, **35**, W169–W175.

Huang, S. *et al.* (2014) A novel model to combine clinical and pathway-based transcriptomic information for the prognosis prediction of breast cancer. *PLOS Comput. Biol.*, **10**, e1003851.

Huang, Z. *et al.* (2020) Deep learning-based cancer survival prognosis from RNA-seq data: approaches and evaluations. *BMC Med. Genomics*, **13** Supplement 5, 41.

Issa, N.T. *et al.* (2021) Machine and deep learning approaches for cancer drug repurposing. *Semin. Cancer Biol.*, **68**, 132–142.

Kanehisa, M. and Goto, S. (2000) KEGG: kyoto encyclopedia of genes and genomes. *Nucleic Acids Res.*, **28**, 27–30.

Kanehisa, M. *et al.* (2016) KEGG as a reference resource for gene and protein annotation. *Nucleic Acids Res.*, **44**, D457–D462.

Katzman, J.L. *et al.* (2018) DeepSurv: personalized treatment recommender system using a Cox proportional hazards deep neural network. *BMC Med. Res. Methodol.*, **18**, 24.

Kipf, T.N. and Welling, M. Semi-supervised classification with graph convolutional networks. International Conference on Learning Representations (2017).

Li, H. *et al.* (2018) The toxic effects of aflatoxin b1 and aflatoxin M1 on kidney through regulating L-proline and downstream apoptosis. *BioMed Res. Int.*, **2018**, 9074861.

Liberzon, A. *et al.* (2015) The Molecular Signatures Database (MSigDB) hallmark gene set collection. *Cell Syst.*, **1**, 417–425.

Liu, J. *et al.* (2018) An integrated TCGA pan-cancer clinical data resource to drive high-quality survival outcome analytics. *Cell*, **173**, 400–416.e11.

Marchese, S. *et al.* (2018) Aflatoxin b1 and M1: biological properties and their involvement in cancer development. *Toxins (Basel)*, **10**.

Newman, A.M. *et al.* (2015) Robust enumeration of cell subsets from tissue expression profiles. *Nat. Methods*, **12**, 453–457.

Pass, H.I. *et al.* (2005) Asbestos exposure, pleural mesothelioma, and serum osteopontin levels. *N. Engl. J. Med.*, **353**, 1564–1573.

Paszke, A. *et al.* (2019) PyTorch: an imperative style, high-performance deep learning library. In, *Adv. Neural Inf. Process. Syst.*, **32**, 8024–8035.

Pavageau, M. *et al.* (2021) DeepOS: pan-cancer prognosis estimation from RNA-sequencing data. *medRxiv*.

Pedregosa, F. *et al.* (2011) Scikit – learn: machine Learning in Python. *J. Mach. Learn. Res.*, **12**, 2825–2830.

Poirion, O.B. *et al.* (2021) DeepProg: an ensemble of deep-learning and machine-learning models for prognosis prediction using multi-omics data. *Genome Med.*, **13**, 112.

Ramirez, R. *et al.* (2020) Classification of cancer types using graph convolutional neural networks. *Front. Phys.*, **8**.

Ramirez, R. *et al.* (2021) Prediction and interpretation of cancer survival using graph convolution neural networks. *Methods*, **192**, 120–130.

, V., Zanirato, G. and Aldieri, E. (2021) The epithelial-to-mesenchymal transition (EMT) in the development and metastasis of malignant pleural mesothelioma. *Int. J. Mol. Sci.*, **22**.

Schelch, K. *et al.* (2018) FGF2 and EGF induce epithelial-mesenchymal transition in malignant pleural mesothelioma cells via a MAPKinase/MMP1 signal. *Carcinogenesis*, **39**, 534–545.

Shannon, P. *et al.* (2003) Cytoscape: a software environment for integrated models of biomolecular interaction networks. *Genome Res.*, **13**, 2498–2504.

Sundararajan, M. *et al* Axiomatic attribution for deep networks. In, *Proceedings of the 34th international conference on* machine learning sydney, NSW, Australia (2017). p. 3319–3328.

Tajima, K. *et al.* (2010) Osteopontin-mediated enhanced hyaluronan binding induces multidrug resistance in mesothelioma cells. *Oncogene*, **29**, 1941–1951.

Xu, K. *et al* How powerful are graph neural networks? In, International Conference on Learning Representations (2019).

Yang, B. *et al.* (2014) Embedding entities and relations for learning and inference in knowledge bases. *Clin. Orthop. Relat. Res.*, abs./**1412.6575**.

Zaheer, M. *et al.* (2017) Deep sets. In, *Adv. Neural Inf. Process. Syst.*, **30**.

Zheng, X., Amos, C.I. and Frost, H.R. (2020) Comparison of pathway and gene-level models for cancer prognosis prediction. *BMC Bioinformatics*, **21**, 76.

Zheng, X., Amos, C.I. and Frost, H.R. (2021) Pan-cancer evaluation of gene expression and somatic alteration data for cancer prognosis prediction. *BMC Cancer*, **21**, 1053.

Zhou, B. and Gao, S. (2021) Pan-cancer analysis of FURIN as a potential prognostic and immunological biomarker. *Front. Mol. Biosci.*, **8**, 648402.

# Supplementary Materials

## Supplementary Figures

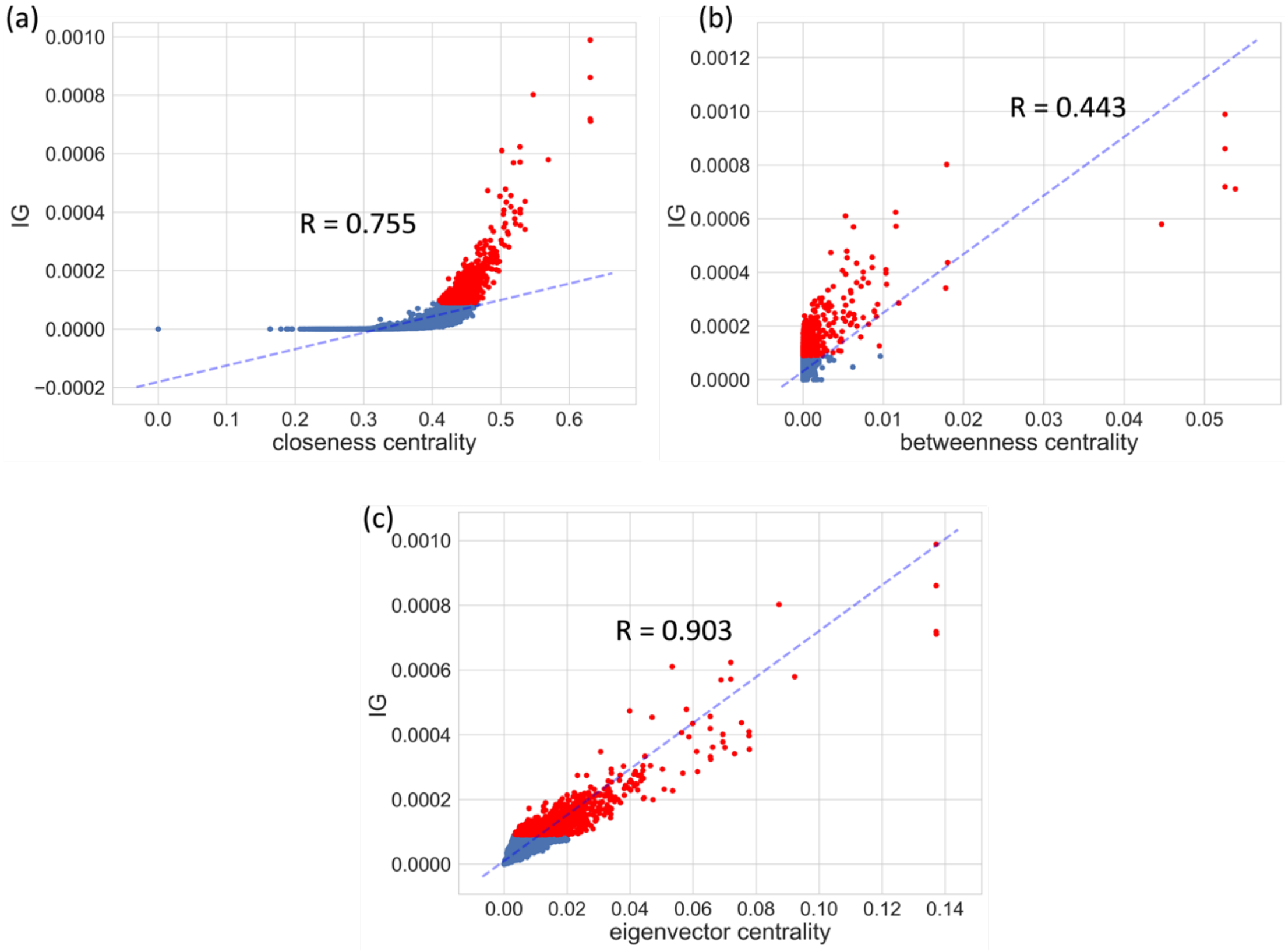

**Figure S1. The graph node IG scatter plot for three centralities.** a) The horizontal axis is closeness centrality. b) The horizontal axis is betweenness centrality. c) The horizontal axis is eigenvector centrality.

## Supplementary Tables

**Table 1**. Sample details. Each year column represents verification year. A column "death" represents the number of patients who died within each verification year. A column "survival" represents the number of patients who have survived for more than each year. A column "total" represents the number of patients added in the two columns. Abbreviations of cancer types are based on The Cancer Genome Atlas (TCGA) notation.

| | year 1 | | | year 2 | | | year 3 | | | year 4 | | | year 5 | | |
|---|---|---|---|---|---|---|---|---|---|---|---|---|---|---|---|
| cancer type | total | death | survival | total | death | survival | total | death | survival | total | death | survival | total | death | survival |
| **ACC** | 74 | 5 | 69 | 69 | 16 | 53 | 57 | 18 | 39 | 47 | 22 | 25 | 44 | 25 | 19 |
| **BLCA** | 385 | 86 | 299 | 294 | 156 | 138 | 259 | 176 | 83 | 243 | 182 | 61 | 227 | 186 | 41 |
| **BRCA** | 1037 | 23 | 1014 | 710 | 47 | 663 | 554 | 88 | 466 | 448 | 107 | 341 | 377 | 131 | 246 |
| **CESC** | 234 | 20 | 214 | 168 | 44 | 124 | 131 | 57 | 74 | 108 | 64 | 44 | 91 | 64 | 27 |
| **CHOL** | 41 | 9 | 32 | 35 | 17 | 18 | 33 | 19 | 14 | 28 | 21 | 7 | 24 | 21 | 3 |
| **COAD** | 467 | 51 | 416 | 336 | 78 | 258 | 227 | 91 | 136 | 166 | 103 | 63 | 152 | 109 | 43 |
| **DLBC** | 37 | 3 | 34 | 32 | 6 | 26 | 20 | 6 | 14 | 16 | 7 | 9 | 12 | 7 | 5 |
| **ESCA** | 164 | 42 | 122 | 106 | 64 | 42 | 90 | 73 | 17 | 88 | 79 | 9 | 84 | 81 | 3 |
| **GBM** | 147 | 66 | 81 | 140 | 115 | 25 | 136 | 128 | 8 | 136 | 133 | 3 | 136 | 134 | 2 |
| **HNSC** | 498 | 100 | 398 | 402 | 180 | 222 | 341 | 205 | 136 | 307 | 213 | 94 | 274 | 225 | 49 |
| **KICH** | 74 | 1 | 73 | 69 | 3 | 66 | 62 | 8 | 54 | 57 | 9 | 48 | 52 | 11 | 41 |
| **KIRC** | 551 | 58 | 493 | 494 | 96 | 398 | 445 | 129 | 316 | 393 | 150 | 243 | 340 | 170 | 170 |
| **KIRP** | 270 | 15 | 255 | 192 | 28 | 164 | 139 | 31 | 108 | 120 | 36 | 84 | 103 | 45 | 58 |
| **LAML** | 98 | 35 | 63 | 84 | 45 | 39 | 79 | 51 | 28 | 71 | 53 | 18 | 63 | 55 | 8 |
| **LGG** | 417 | 26 | 391 | 294 | 57 | 237 | 229 | 79 | 150 | 174 | 97 | 77 | 158 | 105 | 53 |

| | year 1 | | | year 2 | | | year 3 | | | year 4 | | | year 5 | | |
|---|---|---|---|---|---|---|---|---|---|---|---|---|---|---|---|
| cancer type | total | death | survival | total | death | survival | total | death | survival | total | death | survival | total | death | survival |
| **LIHC** | 367 | 68 | 299 | 275 | 109 | 166 | 237 | 127 | 110 | 217 | 140 | 77 | 200 | 150 | 50 |
| **LUAD** | 525 | 67 | 458 | 387 | 121 | 266 | 320 | 161 | 160 | 267 | 187 | 80 | 254 | 203 | 51 |
| **LUSC** | 461 | 89 | 372 | 382 | 146 | 236 | 325 | 178 | 147 | 301 | 195 | 106 | 285 | 208 | 77 |
| **MESO** | 83 | 27 | 56 | 81 | 51 | 30 | 80 | 66 | 14 | 77 | 69 | 8 | 76 | 71 | 5 |
| **OV** | 368 | 39 | 329 | 347 | 91 | 256 | 319 | 134 | 185 | 305 | 186 | 119 | 298 | 219 | 79 |
| **PAAD** | 159 | 42 | 117 | 122 | 86 | 36 | 110 | 90 | 20 | 103 | 92 | 11 | 101 | 93 | 8 |
| **PCPG** | 137 | 5 | 132 | 99 | 6 | 93 | 63 | 7 | 56 | 43 | 7 | 36 | 35 | 7 | 28 |
| **PRAD** | 472 | 1 | 471 | 353 | 3 | 350 | 228 | 5 | 223 | 138 | 6 | 132 | 92 | 6 | 86 |
| **READ** | 141 | 8 | 133 | 94 | 15 | 79 | 64 | 17 | 47 | 41 | 26 | 15 | 34 | 29 | 5 |
| **SARC** | 240 | 27 | 213 | 202 | 54 | 148 | 172 | 70 | 102 | 150 | 79 | 71 | 135 | 87 | 48 |
| **SKCM** | 356 | 28 | 328 | 300 | 77 | 223 | 272 | 112 | 160 | 156 | 131 | 125 | 232 | 148 | 84 |
| **STAD** | 364 | 89 | 275 | 211 | 159 | 52 | 211 | 159 | 52 | 186 | 162 | 24 | 180 | 165 | 15 |
| **TGCT** | 94 | 1 | 93 | 70 | 3 | 67 | 50 | 3 | 47 | 36 | 3 | 33 | 27 | 3 | 24 |
| **THCA** | 487 | 3 | 484 | 342 | 5 | 337 | 223 | 11 | 212 | 153 | 12 | 141 | 114 | 18 | 96 |
| **THYM** | 106 | 2 | 104 | 84 | 3 | 81 | 62 | 5 | 57 | 51 | 5 | 46 | 32 | 6 | 26 |
| **UCEC** | 507 | 24 | 483 | 385 | 49 | 336 | 292 | 74 | 218 | 240 | 84 | 156 | 197 | 87 | 110 |
| **UCS** | 53 | 11 | 42 | 49 | 25 | 24 | 41 | 30 | 11 | 41 | 30 | 11 | 38 | 32 | 6 |
| **UVM** | 69 | 5 | 64 | 58 | 14 | 44 | 43 | 17 | 26 | 28 | 22 | 6 | 26 | 23 | 3 |

## Abbreviations of cancer types

ACC: Adrenocortical carcinoma.

BLCA: Bladder Urothelial Carcinoma

BRCA: Breast invasive carcinoma

CESC: Cervical squamous cell carcinoma and endocervical adenocarcinoma

CHOL: Cholangiocarcinoma

COAD: Colon adenocarcinoma

DLBC: Lymphoid Neoplasm Diffuse Large B-cell Lymphoma

ESCA: Esophageal carcinoma

GBM: Glioblastoma multiforme

HNSC: Head and Neck squamous cell carcinoma

KICH: Kidney Chromophobe

KIRC: Kidney renal clear cell carcinoma

KIRP: Kidney renal papillary cell carcinoma

LAML: Acute Myeloid Leukemia

LGG: Brain Lower Grade Glioma

LIHC: Liver hepatocellular carcinoma

LUAD: Lung adenocarcinoma

LUSC: Lung squamous cell carcinoma

MESO: Miscellaneous

OV: Ovarian serous cystadenocarcinoma

PAAD: Pancreatic adenocarcinoma

PCPG: Pheochromocytoma and Paraganglioma

PRAD: Prostate adenocarcinoma

READ: Rectum adenocarcinoma

SARC: Sarcoma

SKCM: Skin Cutaneous Melanoma

STAD: Stomach adenocarcinoma

TGCT: Testicular Germ Cell Tumors

THCA: Thyroid carcinoma

THYM: Thymoma

UCEC: Uterine Corpus Endometrial Carcinoma

UCS: Uterine Carcinosarcoma

UVM: Uveal Melanoma

**Table. 2.** The enrichment analysis result of the Top 300/500/1000IG node.

(a) The enrichment analysis result of the Top 300 IG node.

| Rank | KEGG Term | Count | P-value |
|---|---|---|---|
| 1 | Pathways in cancer | 109 | 7.85E-51 |
| 2 | Hepatitis B | 59 | 3.18E-41 |
| 3 | Prostate cancer | 48 | 7.76E-41 |
| 4 | Kaposi sarcoma-associated herpesvirus infection | 61 | 2.00E-38 |
| 5 | Lipid and atherosclerosis | 60 | 1.61E-34 |
| 6 | AGE-RAGE signaling pathway in diabetic complications | 43 | 2.77E-33 |
| 7 | Proteoglycans in cancer | 54 | 1.41E-29 |
| 8 | Human cytomegalovirus infection | 55 | 2.25E-28 |
| 9 | EGFR tyrosine kinase inhibitor resistance | 35 | 1.651E-27 |
| 10 | Glioma | 34 | 4.14E-27 |

(b) The enrichment analysis result of the Top 500 IG node.

| Rank | KEGG Term | Count | P-value |
|---|---|---|---|
| 1 | Pathways in cancer | 162 | 3.10E-69 |
| 2 | Hepatitis B | 72 | 1.09E-41 |
| 3 | Prostate cancer | 57 | 2.10E-41 |
| 4 | Lipid and atherosclerosis | 79 | 1.23E-38 |
| 5 | Kaposi sarcoma-associated herpesvirus infection | 74 | 2.10E-37 |
| 6 | AGE-RAGE signaling pathway in diabetic complications | 54 | 1.37E-36 |
| 7 | PI3K-Akt signaling pathway | 94 | 4.24E-33 |
| 8 | EGFR tyrosine kinase inhibitor resistance | 46 | 4.821E-33 |
| 9 | Proteoglycans in cancer | 71 | 1.00E-33 |
| 10 | Apoptosis | 56 | 3.54E-30 |

(c) The enrichment analysis result of the Top 1,000 IG node

| Rank | KEGG Term | Count | P-value |
|---|---|---|---|
| 1 | Pathways in cancer | 230 | 1.21E-74 |
| 2 | Hepatitis B | 94 | 1.13E-42 |
| 3 | Prostate cancer | 71 | 5.12E-42 |
| 4 | Lipid and atherosclerosis | 108 | 3.78E-41 |
| 5 | EGFR tyrosine kinase inhibitor resistance | 62 | 1.15E-39 |
| 6 | Proteoglycans in cancer | 101 | 1.70E-37 |
| 7 | PI3K-Akt signaling pathway | 138 | 2.93E-37 |
| 8 | AGE-RAGE signaling pathway in diabetic complications | 67 | 7.93E-36 |
| 9 | Kaposi sarcoma-associated herpesvirus infection | 96 | 8.09E-36 |
| 10 | Human T-cell leukemia virus 1 infection | 103 | 2.33E-35 |

**Table. 3.** Graph node lists with high IG values for four cancer types.

| Cancer type | Gene | P-value |
|---|---|---|
| ACC | CHEBI: 16842 | 4.43E-02 |
| | SF1 | 3.04E-02 |
| | CHEBI: 44185 | 3.74E-02 |
| | CSTA | 4.96E-02 |
| | GCH1 | 4.67E-02 |
| | ACSL3 | 3.06E-02 |
| KIRP | CHEBI: 2504 | 4.70E-02 |
| | SNRNP70 | 7.77E-03 |
| | PLEKHA7 | 2.85E-02 |
| | ACTG1 | 3.36E-02 |
| | H2AFX | 4.10E-02 |
| | TARDBP | 1.72E-02 |
| | CBX5 | 3.70E-02 |
| | DDIT3 | 2.85E-02 |
| | RPL31 | 1.21E-02 |
| | EIF3E | 1.31E-02 |
| | SMARCC1 | 3.47E-02 |
| | SLC31A1 | 2.21E-02 |
| | NUMA1 | 4.64E-02 |
| | ADD3 | 1.16E-02 |
| | TKT | 4.75E-02 |
| | NKX2-1 | 2.41E-02 |
| KIRP | PEA15 | 2.99E-02 |
| LGG | HSP90AA1 | 4.09E-03 |
| | AR | 3.24E-03 |
| | EFTUD2 | 1.38E-02 |
| | NFKBIA | 1.85E-02 |
| | DDX17 | 2.90E-02 |
| | STAT1 | 4.84E-02 |
| | SGTA | 4.61E-02 |
| | SRRM2 | 2.03E-02 |
| | CHEBI: 27568 | 3.99E-02 |
| | UBE2I | 1.59E-02 |
| | ITGB1 | 2.60E-02 |
| | MAP2K6 | 7.47E-03 |
| | RPS8 | 2.74E-02 |
| | FYN | 1.42E-02 |
| | ADRB2 | 3.34E-02 |
| | RPA1 | 2.13E-02 |
| | STUB1 | 3.60E-03 |
| | DDIT4 | 4.66E-03 |
| | PRKCB | 3.36E-02 |
| | SETD1A | 2.55E-02 |
| | NAPA | 3.16E-02 |
| | FOXP1 | 3.57E-02 |
| | CGB5 | 2.65E-02 |
| LGG | CITED2 | 1.06E-02 |
| | MSMO1 | 4.23E-02 |
| | ACTN4 | 3.30E-04 |
| | SDC1 | 1.34E-02 |
| | CDKN2B | 3.57E-02 |
| | PDIA6 | 3.46E-02 |
| | SEC24D | 8.28E-03 |
| | ZBTB20 | 1.68E-02 |
| | FHL2 | 2.09E-02 |
| | CCT6A | 4.63E-04 |
| | GPAT3 | 2.66E-02 |
| MESO | JUN | 2.66E-02 |
| | CHEBI: 91108 | 1.35E-03 |
| | CHEBI: 30614 | 3.38E-02 |
| | AR | 6.32E-04 |
| | VCP | 4.23E-02 |
| | HSPA1B | 4.51E-02 |
| | XRCC6 | 4.81E-02 |
| | SLBP | 1.48E-02 |
| | SGTA | 1.09E-02 |
| | SRRM2 | 3.64E-02 |
| | ITGA4 | 4.86E-03 |
| | CCL2 | 4.97E-02 |

| Cancer type | Gene | P-value |
|---|---|---|
| | CHEBI: 6820 | 3.77E-02 |
| | PLK1 | 1.80E-02 |
| | POU2F1 | 5.37E-03 |
| | MCM4 | 4.09E-02 |
| | MAP2K6 | 3.35E-02 |
| | IGF1 | 2.30E-02 |
| | LEP | 3.86E-02 |
| | TNFRSF1A | 4.91E-02 |
| | HIST1H2BE | 3.46E-02 |
| | EPB41L2 | 2.45E-02 |
| | DDIT4 | 1.59E-02 |
| | SKP1 | 1.53E-02 |
| | BMP2 | 4.15E-02 |
| | PTBP1 | 4.72E-03 |

| Cancer type | Gene | P-value |
|---|---|---|
| | SETD1A | 3.67E-02 |
| | FGF18 | 4.71E-02 |
| | SULF2 | 3.16E-02 |
| | MAP2K3 | 4.56E-02 |
| | SLC4A7 | 3.75E-02 |
| | KDR | 2.49E-02 |
| | FOXP1 | 1.63E-02 |
| | HIST2H2AC | 3.90E-02 |
| | PPP1CB | 3.75E-02 |
| | DHX15 | 4.97E-02 |
| | RPS13 | 4.23E-02 |
| | TUBA1C | 3.52E-02 |
| | NOTCH1 | 2.10E-02 |
| | NCOA7 | 3.70E-02 |
| | RPS3A | 1.55E-02 |

| Cancer type | Gene | P -value |
|---|---|---|
| | CTNNA1 | 1.76E-02 |
| | INSIG1 | 3.83E-02 |
| | ACTN4 | 3.31E-02 |
| | MAP4 | 2.60E-02 |
| | DEK | 2.30E-02 |
| | PPP1CC | 3.58E-02 |
| | SRF | 7.83E-03 |
| | PDIA6 | 2.19E-02 |
| | NTN4 | 4.70E-02 |
| | DST | 3.24E-03 |
| | EIF3B | 1.31E-02 |
| | CYP19A1 | 1.38E-02 |
| | ABCG1 | 4.11E-02 |
| | NEDD9 | 4.00E-02 |
| | ZIC1 | 3.42E-02 |

**Table. 4.** Gene lists with high IG values for their expression data.

| Cancer type | Gene | P -value |
|---|---|---|
| ACC | WNT2 | 8.53E-05 |
| | GRB7 | 5.69E-10 |
| | FETUB | 9.98E-03 |
| | TKT | 4.83E-03 |
| | CLDN6 | 2.55E-02 |
| | SLC31A1 | 3.99E-02 |
| | PEX14 | 8.05E-07 |
| | ASGR2 | 1.18E-03 |
| | GYPE | 5.04E-05 |
| | CHST3 | 5.13E-05 |
| | ALDH3B1 | 5.03E-06 |
| | ITGA10 | 2.68E-02 |
| | SHROOM2 | 3.06E-02 |
| | CRYBB1 | 1.88E-04 |
| | NFE2 | 6.04E-04 |
| | SDCCAG8 | 1.22E-02 |
| | THEG | 9.39E-04 |
| | ATP5F1D | 3.20E-04 |
| | SCRN1 | 1.90E-04 |
| | NOL3 | 3.24E-02 |
| | IL18BP | 4.28E-04 |
| | TSC22D1 | 9.91E-03 |
| | ARL2BP | 2.39E-02 |
| | PIAS1 | 4.09E-02 |
| | DACH1 | 1.43E-02 |
| ACC | NDUFA7 | 3.26E-07 |
| | SPRR3 | 4.02E-04 |
| | DKK 4.00 | 1.88E-07 |
| | CCR6 | 7.91E-09 |
| | C1orf54 | 2.66E-10 |
| | RXRG | 6.25E-03 |
| | ERBB3 | 2.83E-02 |
| | PIAS3 | 1.44E-02 |
| | ZBTB32 | 3.90E-02 |
| | PCNA | 9.98E-03 |
| KIRP | CYC1 | 4.01E-03 |
| | CHRNA5 | 2.68E-03 |
| | MRPL9 | 9.28E-03 |
| | STBD1 | 2.09E-09 |
| | CASQ2 | 3.22E-04 |
| | VCPIP1 | 1.33E-04 |
| | DST | 7.96E-06 |
| | CLDN6 | 5.67E-03 |
| | ADD2 | 9.60E-06 |
| | CELSR2 | 3.77E-02 |
| | CHST3 | 4.78E-04 |
| | ITGA10 | 4.15E-04 |
| | CRYBB1 | 2.97E-04 |
| | COL4A1 | 7.80E-03 |
| | FMO1 | 7.48E-03 |
| KIRP | COPB2 | 1.93E-02 |
| | NEDD9 | 1.18E-02 |
| | PLAGL1 | 4.37E-02 |
| | SDCCAG8 | 2.62E-04 |
| | TNIP2 | 6.00E-03 |
| | OPN3 | 4.87E-02 |
| | ATP5F1D | 3.76E-03 |
| | FGF13 | 3.87E-03 |
| | CASP10 | 4.04E-02 |
| | GPX8 | 1.22E-03 |
| | SCRN1 | 1.70E-08 |
| | TLX1 | 2.41E-02 |
| | DNAJA1 | 1.46E-04 |
| | DACH1 | 1.14E-04 |
| | RHOA | 2.49E-04 |
| | SPRR3 | 1.55E-03 |
| | CCR6 | 2.39E-09 |
| | RXRG | 4.80E-03 |
| | CRYBG1 | 8.20E-03 |
| | HELLS | 1.69E-02 |
| | PIAS3 | 3.40E-03 |
| | CDKN1C | 3.20E-04 |
| | PCNA | 1.37E-03 |

| Cancer type | Gene | P -value |
|---|---|---|
| LGG | WNT2 | 1.42E-25 |
| | PSMD13 | 3.66E-07 |
| | TUBA4A | 1.72E-07 |
| | RYR1 | 2.64E-21 |
| | OAZ3 | 7.05E-06 |
| | HLA-E | 8.96E-06 |
| | ITIH3 | 9.64E-06 |
| | FETUB | 1.95E-02 |
| | BCAN | 1.99E-05 |
| | MRPL9 | 1.24E-07 |
| | TSPAN4 | 1.91E-04 |
| | HSPA9 | 5.47E-06 |
| | MAP7 | 1.77E-09 |
| | DGKH | 2.58E-07 |
| | CASQ2 | 3.74E-04 |
| | CD6 | 5.24E-06 |
| | TTPA | 9.33E-05 |
| | PTPRE | 2.68E-06 |
| | SH3BGR | 9.63E-12 |
| | IGSF3 | 8.41E-05 |
| | EEF1E1 | 6.29E-04 |
| | DST | 2.21E-33 |
| | PGLS | 2.90E-04 |
| | SLC31A1 | 4.69E-09 |
| | ASGR2 | 5.63E-09 |
| | ASGR2 | 5.63E-09 |

| Cancer type | Gene | P -value |
|---|---|---|
| LGG | AQP7 | 4.44E-04 |
| | CELSR2 | 8.66E-03 |
| | CHST3 | 7.32E-15 |
| | ALDH3B1 | 1.25E-02 |
| | ATP5F1B | 8.24E-06 |
| | GNPAT | 3.87E-04 |
| | AEN | 4.05E-04 |
| | CHST15 | 1.51E-04 |
| | GMPR2 | 1.03E-06 |
| | CRYBB1 | 1.06E-04 |
| | COL4A1 | 2.26E-02 |
| | ADRM1 | 4.81E-06 |
| | CADM1 | 4.87E-02 |
| | SLC30A3 | 3.84E-09 |
| | ZFP36 | 4.58E-06 |
| | INPP4B | 3.50E-03 |
| | RAB31 | 4.29E-16 |
| | IMMT | 4.88E-06 |
| | STIL | 8.73E-12 |
| | POLE | 1.50E-06 |
| | COPB2 | 1.53E-02 |
| | VIP | 1.64E-07 |
| | PLAGL1 | 2.41E-02 |
| | SDCCAG8 | 4.57E-02 |
| | THEG | 8.59E-05 |
| | TNIP2 | 1.24E-23 |

| Cancer type | Gene | P -value |
|---|---|---|
| LGG | OPN3 | 1.16E-04 |
| | IGFBP5 | 3.19E-05 |
| | ATP5F1D | 2.08E-03 |
| | CASP10 | 8.65E-04 |
| | SCRN1 | 1.73E-13 |
| | RRP9 | 2.87E-08 |
| | PSMA6 | 3.65E-07 |
| | RAD51 | 2.44E-04 |
| | PTGER2 | 3.35E-02 |
| | NOL3 | 4.16E-09 |
| | PUS1 | 1.41E-05 |
| | CISH | 3.70E-06 |
| | EFEMP2 | 7.98E-03 |
| | TLX1 | 1.87E-05 |
| | ARL2BP | 8.97E-06 |
| | PTCD3 | 2.10E-02 |
| | RHOA | 4.64E-39 |
| | SPRR3 | 5.80E-17 |
| | GAD1 | 2.22E-03 |
| | TRO | 3.52E-02 |
| | CMKLR1 | 1.14E-04 |
| | VIM | 1.40E-02 |
| | CCR6 | 9.22E-04 |
| | RXRG | 1.83E-07 |
| | CRYBG1 | 1.68E-20 |
| | HELLS | 1.78E-04 |

| Cancer type | Gene | P -value |
|---|---|---|
| LGG | SULT1B1 | 9.00E-04 |
| | PIAS3 | 8.62E-04 |
| | CDKN1C | 5.68E-18 |
| | ZBTB32 | 1.38E-04 |
| MESO | TKT | 2.85E-02 |
| | DST | 4.58E-02 |
| | CCNB2 | 5.19E-11 |
| | CELSR2 | 6.88E-03 |

| Cancer type | Gene | P -value |
|---|---|---|
| MESO | ITGA10 | 3.82E-02 |
| | COL4A1 | 9.98E-03 |
| | IGFBP5 | 2.25E-02 |
| | FGF13 | 3.44E-02 |
| | CASP10 | 1.35E-02 |
| | PSMA6 | 1.78E-02 |
| | DNAJA1 | 5.93E-04 |
| | RHOA | 4.47E-05 |

| Cancer type | Gene | P -value |
|---|---|---|
| MESO | SPRR3 | 6.37E-03 |
| | MZB1 | 8.99E-03 |
| | CCR6 | 3.72E-04 |
| | C1orf54 | 4.14E-23 |
| | SULT1B1 | 7.09E-03 |
| | PCNA | 2.89E-02 |